\documentclass[letterpaper,nomarginnotes,nonarrowgutter]{jpaper}
\usepackage[sort,compress]{cite}
\usepackage{amsmath,amssymb,amsfonts}
\usepackage{algorithmic}
\usepackage{graphicx}
\usepackage{textcomp}
\usepackage{xcolor}
\usepackage{booktabs}
\usepackage{multirow}
\usepackage{caption}
\usepackage{url} 
\usepackage{setspace}
\usepackage{pifont}
\usepackage{footmisc}
\usepackage{balance}
\usepackage{fancyhdr}
\usepackage[bookmarks=true,breaklinks=true,letterpaper=true,colorlinks,citecolor=blue,linkcolor=blue,urlcolor=blue]{hyperref}

\makeatletter
\g@addto@macro{\normalsize}{%
  \setlength{\abovedisplayskip}{4pt plus 0.5pt minus 1pt}
  \setlength{\belowdisplayskip}{3pt plus 0.5pt minus 1pt}
  \setlength{\abovedisplayshortskip}{0pt}
  \setlength{\belowdisplayshortskip}{0pt}
  \setlength{\intextsep}{5pt plus 1pt minus 1pt}
  \setlength{\textfloatsep}{3pt plus 1pt minus 1pt}
  \setlength{\skip\footins}{5pt plus 1pt minus 1pt}
  \setlength{\abovecaptionskip}{2pt plus 0pt minus 0pt}}
\makeatother

\usepackage{titlesec}
\titlespacing\section{0pt}{5pt plus 1pt minus 1pt}{2pt plus 1pt minus 1pt}
\titlespacing\subsection{0pt}{5pt plus 1pt minus 1pt}{2pt plus 1pt minus 1pt}
\titlespacing\subsubsection{0pt}{5pt plus 1pt minus 1pt}{2pt plus 1pt minus 1pt}

\newcommand{\thetitle}{Revisiting DRAM Read Disturbance: Identifying Inconsistencies\\ Between Experimental Characterization and
Device-Level Studies}
\title{\Large{\thetitle}}

\author{
Haocong Luo\quad
{\.I}smail~Emir~Y{\"u}ksel\quad
Ataberk Olgun\quad
A.~Giray~Ya\u{g}l{\i}k\c{c}{\i}\quad
Onur Mutlu\\
ETH Zurich
}

\begin{document}
\definecolor{gfored}{rgb}{0.580, 0.050, 0.211}
\definecolor{ao}{rgb}{0.007, 0.520, 0.867}
\definecolor{yt}{rgb}{0.875, 0.568, 1.000}
\definecolor{moegi}{rgb}{0.357, 0.537, 0.188}
\definecolor{jl}{rgb}{1.0, 0.2, 0.8}
\definecolor{brown(web)}{rgb}{0.65, 0.16, 0.16}
\definecolor{bisque}{rgb}{1.0, 0.89, 0.77}

\newcommand{\om}[1]{\textcolor{red}{#1}}
\newcommand{\hluo}[1]{\textcolor{blue}{#1}}

\newcommand{\exploitingRowHammerAllCitations}[0]{\cite{rowhammer-js,  fournaris2017exploiting, poddebniak2018attacking, tatar2018throwhammer, carre2018openssl, barenghi2018software, zhang2018triggering, bhattacharya2018advanced, google-project-zero, kim2014flipping, rowhammergithub, seaborn2015exploiting, van2016drammer, razavi2016flip, pessl2016drama, xiao2016one, bosman2016dedup, bhattacharya2016curious, burleson2016invited, qiao2016new, brasser2017can, jang2017sgx, aga2017good, mutlu2017rowhammer, tatar2018defeating, gruss2018another, lipp2018nethammer, van2018guardion, frigo2018grand, cojocar2019eccploit,  ji2019pinpoint, mutlu2019rowhammer, hong2019terminal, kwong2020rambleed, frigo2020trrespass, cojocar2020rowhammer, weissman2020jackhammer, zhang2020pthammer, yao2020deephammer, deridder2021smash, hassan2021utrr, jattke2022blacksmith, tol2022toward, kogler2022half, orosa2022spyhammer, zhang2022implicit, liu2022generating, cohen2022hammerscope, zheng2022trojvit, fahr2022frodo, tobah2022spechammer, rakin2022deepsteal, Juffinger2024PressHammer, mutlu2023fundamentally, jattke2024zenhammer, deridder2025posthammer, Kaur2023FlippingBitsLikeaPro, Tol2023DontKnock, Mus2023Jolt, Li2023FPHammer, Baek2025Marionette}}

\newcommand{\hcf}{{HC\textsubscript{First}\ }}
\newcommand{\hcs}{{HC\textsubscript{1$\rightarrow$0Exceeds0$\rightarrow$1}\ }}

\newcommand{\figref}[1]{Figure~\ref{#1}}
\newcommand{\figsref}[1]{Figures~\ref{#1}}
\newcommand{\secref}[1]{§\ref{#1}}
\newcommand{\secsref}[1]{\agycomment{PLEASE USE \\secref MACRO}~\secref{#1}}
\newcommand{\tabref}[1]{Table~\ref{#1}}
\newcommand{\tabsref}[1]{Tables~\ref{#1}}
\newcommand{\obsvref}[1]{Obsv.~\ref{#1}}
\newcommand{\obsvsref}[1]{Obsvs.~\ref{#1}}

\newcounter{char}
\setcounter{char}{0}
\newcommand\characteristicsbox[1]{%
   \refstepcounter{char}
  \vspace{0.2em}
  \noindent
  \begin{tabular}{|p{0.95\linewidth}|}
       \hline
       \textbf{{Characteristic \thechar}.} \emph{{#1}}\\
       \hline 
  \end{tabular}
  \vspace{0.2em}
}

\newcounter{obs}
\setcounter{obs}{0}
\newcommand\observation[1]{%
   \refstepcounter{obs}
  \vspace{0.1em}
  \noindent
  \begin{tabular}{|p{0.95\linewidth}|}
       \hline
       \textbf{{Observation \theobs}.} {{#1}}\\
       \hline 
  \end{tabular}
  \vspace{0.1em}
}

\newcounter{take}
\setcounter{take}{0}
\newcommand\takeaway[1]{%
   \refstepcounter{take}
  \vspace{0.1em}
  \noindent
  \begin{tabular}{|p{0.95\linewidth}|}
       \hline
       \textbf{{Takeaway \thetake}.} {\emph{#1}}\\
       \hline 
  \end{tabular}
  \vspace{0.1em}
}

\setstretch{0.955}
\maketitle

\thispagestyle{plain}
\begin{abstract}
 Modern DRAM is vulnerable to read disturbance (e.g., RowHammer and RowPress) that significantly undermines the robust operation {of the} system. Repeatedly opening and closing a DRAM row (RowHammer) or keeping a DRAM row open for a long period of time (RowPress) induces bitflips in nearby \emph{unaccessed} DRAM rows. Prior {works} on DRAM read disturbance either 1) perform experimental characterization using commercial-off-the-shelf (COTS) DRAM chips to demonstrate the high-level characteristics of the read disturbance bitflips, or 2) perform {device-level simulations} to understand the low-level error mechanisms of the read disturbance bitflips. 
 
 In this paper, we attempt to align and cross-validate the real-chip experimental characterization results and state-of-the-art device-level studies of DRAM read disturbance. To do so, we first identify and extract the key bitflip characteristics of RowHammer and RowPress from the device-level error mechanisms studied in prior works. {Then, we perform experimental characterization on {96} COTS DDR4 DRAM chips that directly match the data and access patterns studied in the device-level works}. Through our experiments, we identify fundamental inconsistencies in the RowHammer and RowPress bitflip directions and access pattern dependence between experimental characterization results and the device-level error mechanisms. 
 
 Based on our results, we hypothesize that either 1) the retention {failure} based DRAM architecture reverse-engineering methodologies do not fully work on modern DDR4 DRAM chips, or 2) existing device-level works do not fully uncover all the major read disturbance error mechanisms. We hope our findings {inspire and} enable future works to build a more fundamental and comprehensive understanding of DRAM read disturbance.
\end{abstract}

\section{Introduction}

Memory isolation is critical to ensure the robust (i.e., safe, secure, and reliable) operation of {modern computing systems}. Accessing a memory address should not have \emph{unintended} side-effects on {data stored in} other \emph{unaccessed} memory addresses. Unfortunately, dynamic random access memory (DRAM){~\cite{dennard1968dram}}, the major main memory technology, suffers from read disturbance {(e.g., RowHammer and RowPress){~\cite{kim2014flipping, kim2020revisiting, orosa2021deeper, yaglikci2022understanding, Luo2023RowPress, Nam2024DRAMScope, mutlu2017rowhammer, mutlu2019rowhammer, mutlu2023fundamentally, Luo2024RowPressTopPicks, luo2024combined, mutlu2023retrospectiveflippingbitsmemory, olgun2024read, olgun2025variable}}}, i.e., repeatedly opening and closing a DRAM row (RowHammer) or keeping a DRAM row open for a long period of time (RowPress) induces bitflips in nearby unaccessed DRAM rows. DRAM read disturbance {is} a critical security vulnerability as attackers can {induce and exploit RowHammer and RowPress bitflips to take over a system or leak private or security-critical data}~\exploitingRowHammerAllCitations{}. Therefore, to make DRAM-based main memory {more robust (i.e., safer, more secure, and more reliable)}, it is important to rigorously study and understand read disturbance mechanisms like RowHammer and RowPress.

Prior works on DRAM read disturbance either 1) perform experimental characterization using commercial-off-the-shelf (COTS) DRAM chips~\cite{kim2014flipping, kim2020revisiting, orosa2021deeper, yaglikci2022understanding, Luo2023RowPress, Nam2024DRAMScope, mutlu2017rowhammer, mutlu2019rowhammer, mutlu2023fundamentally, Luo2024RowPressTopPicks, luo2024combined, mutlu2023retrospectiveflippingbitsmemory, olgun2024read, olgun2025variable} to demonstrate the high-level characteristics of the read disturbance bitflips, or 2) perform device-level simulations to understand the low-level error mechanisms that induce the bitflips~\cite{ryu2017overcoming, yang2019trap, walker2021ondramrowhammer, Zhou2023Double, Zhou2024Unveiling, Jie2024Understanding, Zhou2024Understanding}. Unfortunately, observations from experimental characterization works do not always match the device-level phenomena because the DRAM architecture and array layout is {not visible from} the DRAM chip-level. For example, a DRAM cell can represent a logical ``1'' value by either having $\mathtt{V_{Core}}$ (i.e., a true-cell) or $\mathtt{GND}$ (i.e., an anti-cell) at the {storage} node (vice versa for representing a logical ``0'')~\cite{liu2013experimental}. Thus, observing a ``1'' to ``0'' bitflip through experimental characterization of COTS DRAM chips does not always mean a DRAM cell's storage node voltage drops from $\mathtt{V_{Core}}$.

{\textbf{Our goal}} in this paper is to align and cross-validate the experimental characterization of read disturbance with the low-level fundamental error mechanisms modeled by {device-level simulation works}. We do so by comparing two fundamental properties of the read disturbance bitflips: 1) bitflip direction, and 2) the dependence on DRAM cell data pattern and DRAM row access pattern. To do so, we first identify and extract the key first-order read disturbance error-mechanisms from state-of-the-art device-level works~\cite{ryu2017overcoming, yang2019trap, walker2021ondramrowhammer, Zhou2023Double, Zhou2024Unveiling, Jie2024Understanding, Zhou2024Understanding}, and then perform experimental characterization of DRAM read disturbance on {96 COTS DDR4 DRAM chips from all three major manufacturers (Samsung, SK Hynix, and Micron)} {using DRAM Bender~\cite{olgun2022drambender, hassan2017softmc}} 1) that directly match the data and access patterns studied in device-level works, and 2) apply state-of-the-art DRAM architecture and true- and anti-cell layout reverse engineering techniques based on DRAM cell retention failures~\cite{liu2013experimental, patel2017reaper, khan2014efficacy, khan2016parbor, patel2020beer, Nam2024DRAMScope} to try to match the characterization results to device-level phenomena. 

We identify the following \textbf{key inconsistencies} between experimental characterization and state-of-the-art proposed device-level mechanisms:
\begin{itemize}
    \item For double-sided RowHammer, experimental characterization shows bitflips in {\emph{both}} directions {(``1'' to ``0'' and ``0'' to ``1'')} while device-level mechanisms suggest {\emph{only}} ``1'' to ``0'' bitflips will happen.
    \item For double-sided RowHammer, experimental characterization shows that ``0'' to ``1'' bitflips appear {\emph{first}} while device-level mechanisms suggest the error mechanism to induce ``1'' to ``0'' bitflips is significantly stronger than ``0'' to ``1'' bitflips.
    \item For single-sided RowPress, experimental characterization shows {\emph{overwhelmingly}} ``1'' to ``0'' bitflips while device-level mechanisms suggest {\emph{both}} kinds of bitflips will happen.
\end{itemize}

 Based on our results, we hypothesize that either 1) existing device-level works do not fully uncover all the major read disturbance error mechanisms, or 2) {existing retention-failure-based} DRAM architecture reverse-engineering methodologies do not fully work on modern DDR4 DRAM chips. We hope that our findings enable future works to build a more fundamental and comprehensive understanding of DRAM read disturbance.
 
We make the following contributions in this paper:
\begin{itemize}
    \item To our knowledge, this is the first paper to systematically demonstrate the fundamental inconsistencies of DRAM read disturbance bitflip characteristics between real-chip experimental characterization and proposed device-level error mechanisms.

    \item The observations from our experimental characterization provides insights for future works to build a more fundamental and comprehensive understanding of DRAM read disturbance.

    \item We open source~\cite{vts25github} all our {experimental} infrastructure, code, and data to facilitate future research on DRAM read disturbance.

\end{itemize}

\section{Background}
We provide a brief explanation on 1) the logical organization and key operations of DRAM-based main memory, 2) the physical structure of a DRAM cell, and 3) the DRAM read disturbance phenomenon and the key read disturbance bitflip characteristics modeled by state-of-the-art device-level TCAD simulation works.
\subsection{DRAM-Based Main Memory}

\begin{figure}[h]
    \centering
    \includegraphics[width=1.0\linewidth]{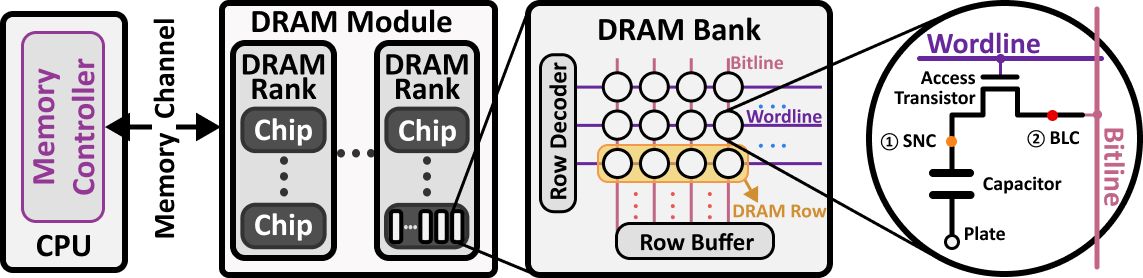}
    \caption{Logical organization of modern DRAM-based main memory. {Reproduced from~\cite{Luo2023RowPress}}.}
    \label{fig:dram_org}
\end{figure}

\figref{fig:dram_org} illustrates the \emph{logical} organization of modern DRAM-based main memory. The CPU's \emph{memory controller} controls a \emph{memory channel} that one or more \emph{DRAM modules} connect to. A module contains one or multiple \emph{DRAM ranks} that share the memory channel. A rank is made up of multiple \emph{DRAM chips} that {operate} in lock-step. Each DRAM chip contains multiple \emph{DRAM banks} that can be accessed independently.

In a DRAM bank, \emph{DRAM cells} are organized in a 2D array addressed by rows (wordlines) and columns (bitlines). A DRAM cell stores {one} bit of information in the form of electric charge in its capacitor. The capacitor is connected to the bitline through the access transistor, controlled by the wordline. The other end of the capacitor is connected to a constant voltage source (plate, usually $\mathtt{V_{Core}}/2$) to reduce the stress of the electric field across the capacitor~\cite{Kraus1989Optimized, keeth2007dram}.

To access DRAM, the memory controller first sends an \texttt{ACT} (activate) command with a row address {that 1) opens a DRAM row in a bank, and 2) transfers the data stored in the DRAM cells to the row buffer ({i.e.,} bitline sense amplifiers, BLSA)}. Second, the memory controller sends {a} \texttt{RD/WR} command with a column address to read or write the desired data in the row buffer. Third, to prepare the DRAM bank for subsequent accesses to other DRAM rows, the memory controller sends a \texttt{PRE} (precharge) command to close the opened DRAM row and reset the row buffer for the next accesses.

Because the BLSA is essentially a cross-coupled pair of inverters, a DRAM cell can {represent either} a logical ``1'' as storing positive charge (i.e., true-cell) or storing negative charge (i.e., anti-cell), depending on which side of the BLSA its bitline is connected to (vice versa for representing a logical ``0'')~\cite{liu2013experimental, patel2017reaper, Nam2024DRAMScope}.
\subsection{Physical DRAM Cell Structure}

\begin{figure}[h]
    \centering
    \includegraphics[width=1.0\linewidth]{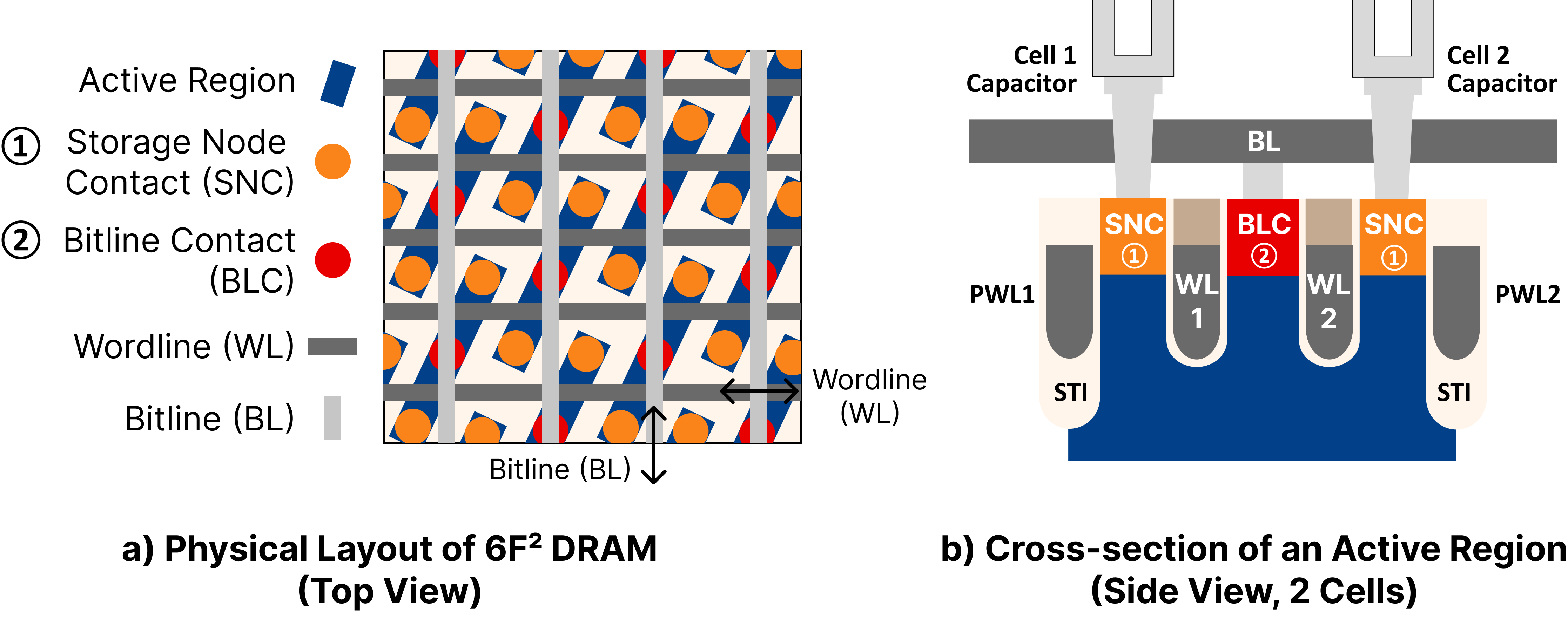}
    \caption{Physical layout and structure of DRAM cell.}
    \label{fig:dram_layout}
\end{figure}

\figref{fig:dram_layout}.a illustrates the physical layout of modern high-density open-bitline $6F^2$ DRAM cell array. {The capacitor of a DRAM cell is connected to an active region (i.e., the area where transistors are formed) through the storage node contact (SNC, \ding{172} in \figref{fig:dram_org} and \figref{fig:dram_layout}).} Two DRAM cells belonging to two neighboring wordlines (rows) are {fabricated} in the same active region, and they share the same bitline contact {(BLC, \ding{173} in \figref{fig:dram_org} and \figref{fig:dram_layout})}. 

\figref{fig:dram_layout}.b shows the physical structure of DRAM cells using a side view of a cross section of an active region containing two cells. For any DRAM cell (e.g., Cell 1 in \figref{fig:dram_layout}.b), two other wordlines are in close proximity that can become potential aggressor rows to induce DRAM read disturbance: 1) the neighboring wordline (NWL) that controls the other DRAM cell in the same active region (i.e., WL2, Cell 2's wordline), and 2) the passing wordline (PWL) in the shallow trench isolation (STI) region that does not form an access transistor in the active region (i.e., PWL1).

\subsection{DRAM Read Disturbance}
DRAM read disturbance is {the} phenomenon that accessing a DRAM row (aggressor row) disturbs the data stored in nearby \emph{unaccessed} rows (victim rows). Prior works~\cite{kim2014flipping, kim2020revisiting, orosa2021deeper, yaglikci2022understanding, Luo2023RowPress, Nam2024DRAMScope, mutlu2017rowhammer, mutlu2019rowhammer, mutlu2023fundamentally, Luo2024RowPressTopPicks, luo2024combined, mutlu2023retrospectiveflippingbitsmemory, olgun2024read, olgun2025variable} demonstrate that DRAM is vulnerable to two types of read disturbance phenomena, RowHammer~\cite{kim2014flipping, kim2020revisiting, orosa2021deeper, yaglikci2022understanding, mutlu2017rowhammer, mutlu2019rowhammer, mutlu2023fundamentally, mutlu2023retrospectiveflippingbitsmemory, olgun2024read, olgun2025variable} and RowPress~\cite{Luo2023RowPress, Nam2024DRAMScope, Luo2024RowPressTopPicks, luo2024combined, olgun2024read, olgun2025variable}.

\noindent\textbf{RowHammer.} RowHammer is a read disturbance phenomenon {where} repeatedly opening (activating) and closing the aggressor DRAM row \emph{many times} (e.g., tens of thousands of times~\cite{kim2020revisiting, orosa2021deeper}) causes bitflips in the victim rows. 

\noindent\textbf{RowPress.} RowPress is a read disturbance phenomenon {where} keeping the aggressor row open for a \emph{long period of time} {causes} bitflips in victim row. Doing so {requires} much fewer aggressor row activations to cause bitflips compared to RowHammer (i.e., orders of magnitude reduction~\cite{Luo2023RowPress, luo2024combined}).

{In the next section, we {1)} summarize the fundamental error mechanisms modeled and studied in state-of-the-art device-level research on DRAM read disturbance~\cite{ryu2017overcoming, yang2019trap, walker2021ondramrowhammer, Zhou2023Double, Zhou2024Unveiling, Jie2024Understanding, Zhou2024Understanding}, and {2)} identify the key read disturbance bitflip characteristics {based on the most fundamental and dominating device-level error mechanisms studied in prior works}.}

\section{Device-Level Mechanisms from Prior Works}
\subsection{Key Error Mechanisms of RowHammer}
When only the neighboring wordline (NWL, i.e., the wordline that shares the same active region with the victim DRAM cell) is the aggressor row in RowHammer, {\emph{trap-assisted electron migration}} is the major error mechanism for RowHammer~\cite{yang2019trap, walker2021ondramrowhammer, Zhou2023Double}. The charge traps near the NWL silicon/gate interface are filled with electrons when NWL is activated. These electrons then get released when NWL is closed, and some of them {migrate} to the victim storage node causing a {``1'' to ``0''} leakage. When only the passing wordline (i.e., PWL) is the aggressor row in RowHammer, the {\emph{passing gate effect}}~\cite{hong2023dsac, Zhou2023Double, Nam2024DRAMScope} pulls electrons away from the storage node of the victim, causing a {``0'' to ``1''} leakage.

For double-sided RowHammer (i.e., {when both} NWL and PWL are activated in an alternating {access pattern}), the trap-assisted electron migration mechanism is \emph{significantly enhanced} because {PWL is open} when NWL is off~\cite{Zhou2023Double}. The electric field from the open PWL significantly enhances the migration of the electrons from traps near NWL to the victim node during the off-phase of the NWL, causing a significantly stronger {``1'' to ``0''} leakage.~\cite{Zhou2023Double} further claims ``0'' to ``1'' bitflips are ``eliminated completely''. Assuming this is indeed the major error mechanism for double-sided RowHammer, then the bitflips that manifest should have the following characteristics.

\characteristicsbox{Double-sided RowHammer should induce only ``1'' to ``0'' bitflips.}

\subsection{Key Error Mechanisms of RowPress}
\label{sec:rowpressmechanism}

When the NWL aggressor is open for a long period of time, more electrons are drawn from the victim storage node to the BLC, causing a stronger {``0'' to ``1''} leakage~\cite{Zhou2024Understanding, Zhou2024Unveiling}. When the PWL aggressor is open for a long period of time, it draws more electrons towards the direction of the victim storage node, causing a stronger {``1'' to ``0''} leakage\cite{Zhou2024Understanding, Zhou2024Unveiling}.

\characteristicsbox{Single-sided RowPress should induce both ``1'' to ``0'' and ``0'' to ``1'' bitflips.}

\section{Experimental Testing Methodology}
\subsection{COTS DDR4 DRAM Testing Infrastructure}
We use DRAM Bender~\cite{olgun2022drambender, safari-drambender} {(built over SoftMC~\cite{hassan2017softmc, softmcgithub})}, an FPGA-based COTS DDR4 DRAM testing infrastructure that gives us fine-grained control of DRAM commands and timings. The infrastructure consists of 1) a host PC that that generates the test program and collects experiment results, 2) an FPGA development board programmed with DRAM Bender that executes the test programs, and 3) the DRAM module under test that is connected to the FPGA board. We also attach a pair of heater pads controlled with a PID-based temperature controller{~\cite{maxwellFT200}} that can maintain and/or increase the temperature of the DRAM.

\subsection{Reverse Engineering DRAM Array Architecture and Layout}
\noindent\textbf{In-DRAM Row Mapping.} We apply the same methodology as prior works~\cite{hassan2021utrr, orosa2021deeper, yaglikci2022understanding, Luo2023RowPress, Nam2024DRAMScope} that {use} single-sided RowHammer to identify the two neighboring rows of an aggressor row.

\noindent\textbf{True- and Anti-Cell Layout.} We apply the same methodology as prior work~\cite{Nam2024DRAMScope} that leverages DRAM retention failure to reverse engineer the layout of true- and anti-cells in DRAM based on the assumption that only a physical ``1'' will experience retention failures~\cite{liu2013experimental, Nam2024DRAMScope}. We find that all the DRAM chips we test from Mfr. S and H have only true-cells, and DRAM chips from Mfr. M have true- and anti-cells interleaved at subarray granularity.\footnote{We also {find that} certain DRAM chips from Mfr. M have both true- and anti-cells within the same row. We do not include them in the study to simplify the testing.}

\subsection{COTS DDR4 DRAM Chips Tested}
We test 12 different {types} of commercial-off-the-shelf (COTS) DDR4 DRAM chips (different die revisions and densities) {from 12 modules (96 chips in total)} from all three major DRAM manufacturers (Mfr. S, H, and M).\footnote{{Our selection of DRAM modules to test is limited by two factors. First, we have access to {only} retail channels, which do not always offer all the DRAM die densities and revisions. Second, to make sure we can align experimental characterization results to device-level mechanisms, we need to ensure we reverse-engineer the DRAM array layout (e.g., row mapping and true- and anti-cell layout) of the DRAM {chips} we test. The array layout of many DRAM {chips} from Mfr. M are difficult to reverse-engineer, so we do not include them in the study to simplify testing.}} Table~\ref{tab:DRAMs} lists all the DRAM chips we test. For every DRAM, we test 2048 rows in bank 1.\footnote{{We assume no significant difference in read disturbance and retention failure characteristics {across} different banks. To save testing time, we test {only} bank 1.}
}

\begin{table}[htbp]
\caption{DRAM Chips Tested}
\label{tab:DRAMs}

\centering

\resizebox{\columnwidth}{!}{%
\begin{tabular}{@{}c|c|cc|c|c|c@{}}
\toprule
{\textbf{Mfr.}} & 
{\textbf{Module Type}} & 
{\textbf{Die Density}} & 
{\textbf{Die Revision}} & 
{\textbf{DQ}} & 
{\textbf{Num. Chips}}& 

\begin{tabular}[c]{@{}c@{}}\textbf{Date Code}\\\textbf{(YYWW)}\end{tabular}
 \\ \midrule
S & UDIMM & 8 Gb  & B & $\times$8 & 8 & 1639 \\
S & UDIMM & 8 Gb  & D & $\times$8 & 8 & 2110 \\
S & UDIMM & 8 Gb  & E & $\times$8 & 8 & 2341 \\
S & UDIMM & 16 Gb & M & $\times$8 & 8 & 2118 \\
S & UDIMM & 16 Gb & A & $\times$8 & 8 & 2319 \\
S & UDIMM & 16 Gb & B & $\times$8 & 8 & 2315 \\
S & UDIMM & 16 Gb & C & $\times$8 & 8 & 2408 \\ \midrule
H & UDIMM & 8 Gb  & C & $\times$8 & 8 & 2120 \\
H & UDIMM & 8 Gb  & D & $\times$8 & 8 & 1938 \\
H & UDIMM & 16 Gb & A & $\times$8 & 8 & 2003 \\
H & UDIMM & 16 Gb & C & $\times$8 & 8 & 2136 \\ \midrule
M & UDIMM & 8 Gb  & E & $\times$8 & 8 & 2402 \\ \bottomrule
\end{tabular}%
}

\end{table}

\section{Experimental Characterization Results}
\subsection{{\hcf with Double-Sided RowHammer}}
\noindent\textbf{Metric.} To align the experimental characterization results with Characteristics 1, we first test the \emph{minimum aggressor row activation (hammer) count to induce at least one bitflip in the victim row}, i.e., {\emph{{HC\textsubscript{First}}}, for both ``0'' to ``1'' and ``1'' to ``0'' bitflips. In double-sided RowHammer, we define \hcf as the per-aggressor-row hammer count.

\noindent\textbf{Methodology.} To find \hcf of a victim row, we sweep the range of possible hammer count values (bounded by the DDR4 DRAM refresh window of 64ms) from 0 to 500K with a step size of 1000 until we find at least one bitflip in the victim row. If we do not find bitflips within 500K hammers, we report we do not find a bitflip. We initialize the victim row with either all-zeros (\texttt{0x00}) or all-ones (\texttt{0xFF}), and initialize both aggressor rows with the opposite data of the victim row. We skip victim rows that do not have two neighboring rows (e.g., at subarray boundaries or are remapped due to repair). We keep the DRAM temperature constant at 50$^\circ$C.

\noindent\textbf{Results.} Figure~\ref{fig:ds_hcf} plots the distribution of the \hcf values (y-axis) with both ``0'' to ``1'' (i.e., the victim row has data \texttt{0x00}) and ``1'' to ``0'' (i.e., the victim row has data \texttt{0xFF}) bitflips across all the victim rows we test for all {types} of DRAM chips we test in box and whiskers plots.\footnote{{The box is bounded by the first quartile (i.e., the median of the first half of the ordered set of data points) and the third quartile (i.e., the median of the second half of the ordered set of data points). The whiskers represent $1.5\times$ the InterQuartile Range (IQR, the distance between the first and third quartiles, i.e., box size). The fliers represent outlier values.\label{figuremethodology}}} A lower \hcf value means it is \emph{easier} to induce bitflips since {doing so} needs fewer aggressor row activations. We make two observations from the results.

\begin{figure}[h]
    \centering
    \includegraphics[width=1.0\linewidth]{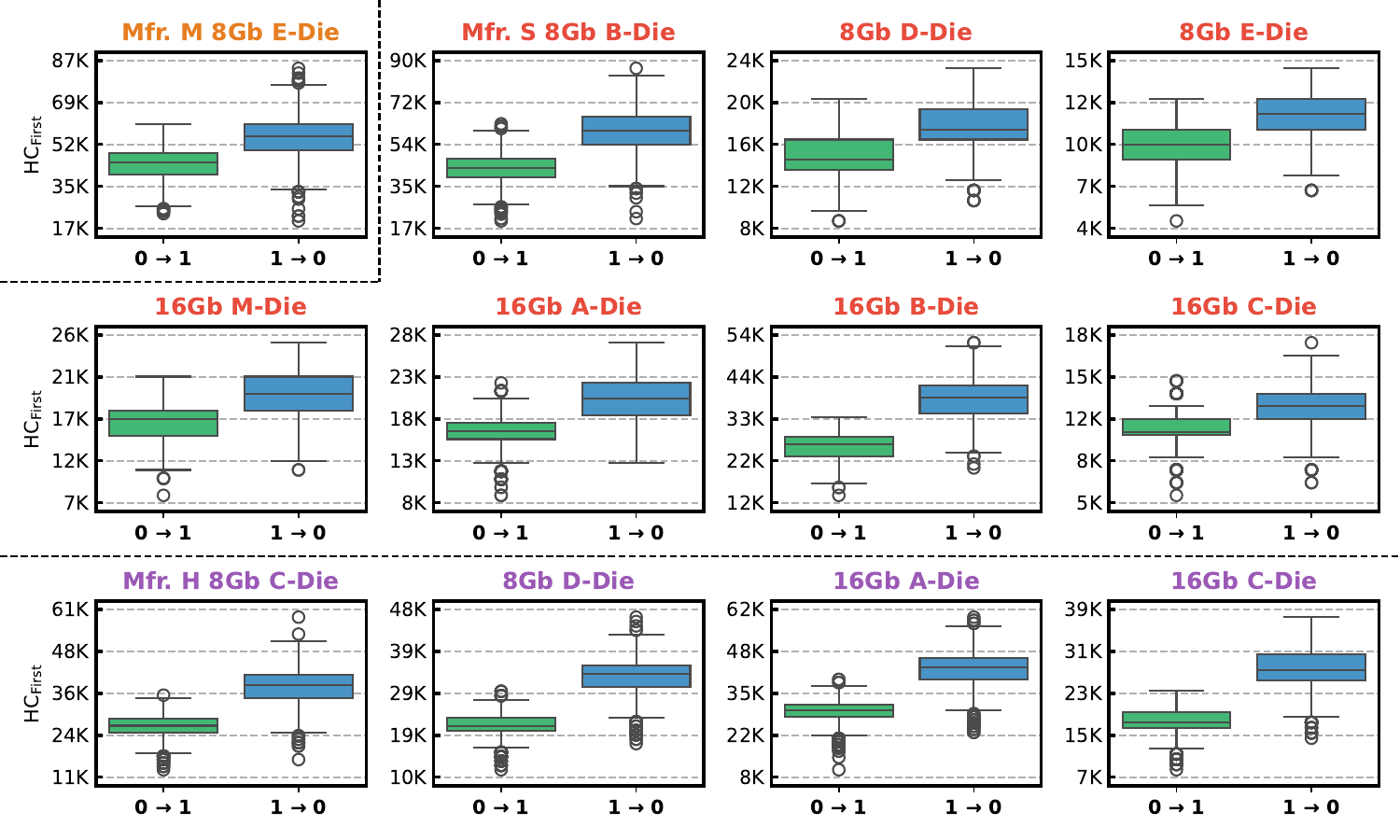}
    \caption{\hcf distribution of ``0'' to ``1'' and ``1'' to ``0'' bitflips (Double-Sided RowHammer).}
    \label{fig:ds_hcf}
\end{figure}

\observation{Double-Sided RowHammer induces both ``0'' to ``1'' and ``1'' to ``0'' bitflips.}

We observe that, for all the DRAM {chips} we test, Double-Sided RowHammer induces both ``0'' to ``1'' and ``1'' to ``0'' bitflips. For the same victim DRAM row, we \emph{always} observe both ``0'' to ``1'' and ``1'' to ``0'' bitflips.

\observation{For Double-Sided RowHammer, the \hcf values of ``0'' to ``1'' bitflips are significantly smaller than that of the ``1'' to ``0'' bitflips.}

We observe that, for all the DRAM {chips} we test, Double-Sided RowHammer induces ``0'' to ``1'' bitflips significantly {more easily} than {it induces} ``1'' to ``0'' bitflips. {Table~\ref{tab:hcf_vals} lists the average \hcf values of both ``0'' to ``1'' and ``1'' to ``0'' bitflips for all the DRAM {chips} we test. {Across} all DRAM {chips} we test, the average \hcf of ``0'' to ``1'' bitflips is {24.7\% smaller than the average} \hcf of ``1'' to ``0'' bitflips. This {implies} that the error mechanism for ``0'' to ``1'' bitflips is stronger than ``1'' to ``0'' bitflips in the \emph{most vulnerable} DRAM cells.}

\begin{table}[h]
\centering
\caption{Average \hcf of ``0'' to ``1'' and ``1'' to ``0'' bitflips (Double-Sided RowHammer).}
\label{tab:hcf_vals}
\resizebox{\columnwidth}{!}{%
\begin{tabular}{@{}c|c|c||cc|c|c@{}}
\toprule
\multirow{2}{*}{\textbf{Mfr.}} &
  \multirow{2}{*}{\textbf{Die Density}} &
  \multirow{2}{*}{\textbf{Die Revision}} &
  \multicolumn{2}{c|}{\textbf{Average \hcf}} &
  \multirow{2}{*}{\textbf{Difference}} &
  \multirow{2}{*}{\textbf{\begin{tabular}[c]{@{}c@{}}Avg. Difference\\ (Geo. Mean)\end{tabular}}} \\
  &      &   & 0 to 1  & 1 to 0  &        &                          \\ \midrule
S & 8 Gb  & B & 43840 & 59368 & 26.2\% & \multirow{12}{*}{24.7\%} \\
S & 8 Gb  & D & 15398 & 18041 & 14.7\% &                          \\
S & 8 Gb  & E & 9684  & 11623 & 16.7\% &                          \\
S & 16 Gb & M & 16732 & 19946 & 16.1\% &                          \\
S & 16 Gb & A & 16981 & 20942 & 18.9\% &                          \\
S & 16 Gb & B & 26415 & 38774 & 31.9\% &                          \\
S & 16 Gb & C & 11355 & 13346 & 14.9\% &                          \\ \cmidrule(r){1-6}
H & 8 Gb  & C & 26500 & 38440 & 31.1\% &                          \\
H & 8 Gb  & D & 22069 & 33489 & 34.1\% &                          \\
H & 16 Gb & A & 29825 & 43326 & 31.2\% &                          \\
H & 16 Gb & C & 18042 & 28041 & 35.7\% &                          \\ \cmidrule(r){1-6}
M & 8 Gb  & E & 44468 & 55605 & 20.0\% &                          \\ \bottomrule

\end{tabular}%
}
\end{table}

We {derive} the following two key takeaways.

\takeaway{Double-Sided RowHammer {involves error mechanisms for inducing} both ``0'' to ``1'' and ``1'' to ``0'' bitflips.}

\takeaway{For Double-Sided RowHammer, the {observed error} mechanism for ``0'' to ``1'' bitflips is stronger than ``1'' to ``0'' bitflips in the most vulnerable DRAM cells {(i.e., those requiring the least number of aggressor row activations to {experience} bitflips)}.}

\subsection{{Bitflip Count with Double-Sided RowHammer}}
\noindent\textbf{Metric.} To further investigate the bitflip characteristics of double-sided RowHammer, we test the maximum number of bitflips that can be induced within the refresh window (i.e., 64ms) for both ``0'' to ``1'' and ``1'' to ``0'' bitflips.

\noindent\textbf{Methodology.} To induce the maximum number of bitflips, we activate each aggressor row 500K times and then count the number of bitflips in the victim row. We initialize the victim row with either all-zeros (\texttt{0x00}) or all-ones (\texttt{0xFF}), and initialize both aggressor rows with the opposite data of the victim row. We skip victim rows that do not have two neighboring rows (e.g., at subarray boundaries or are remapped due to repair). We keep the DRAM temperature constant at 50$^\circ$C.

\noindent\textbf{Results.} Figure~\ref{fig:ds_ber} plots the distribution of the {number of bitflips per victim row} (y-axis) with both ``0'' to ``1'' (i.e., the victim row has data \texttt{0x00}) and ``1'' to ``0'' (i.e., the victim row has data \texttt{0xFF}) bitflips (x-axis) for all {types} of DRAM chips we test in box and whiskers plots.\footref{figuremethodology} We make the following observation from the results.

\begin{figure}[h]
    \centering
    \includegraphics[width=1.0\linewidth]{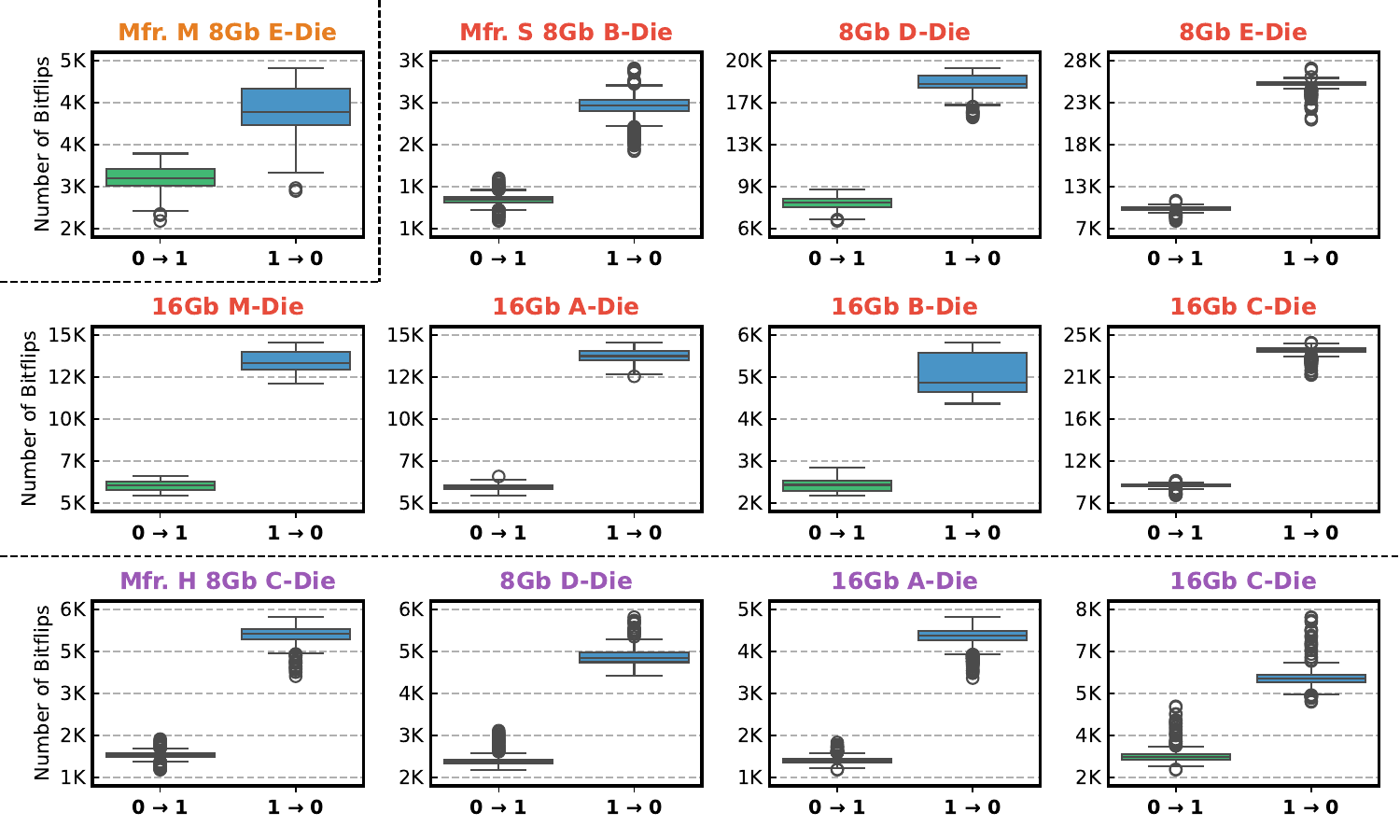}
    \caption{The distribution of the {number of ``0'' to ``1'' and ``1'' to ``0'' bitflips per victim row} (Double-Sided RowHammer).}
    \label{fig:ds_ber}
\end{figure}

\observation{Double-Sided RowHammer induces significantly more ``1'' to ``0'' {bitflips} than ``0'' to ``1'' bitflips at maximum aggressor row activation count.}

We observe that, for all the DRAM {chips} we test, Double-Sided RowHammer induces significantly more ``1'' to ``0'' than ``0'' to ``1'' bitflips at maximum aggressor row activation count. On average, the number of ``1'' to ``0'' bitflips in a victim row is $105.1\%$ {higher than} the number of ``0'' to ``1'' bitflips. This means that when the aggressor rows are hammered for sufficiently high {number of} times, more DRAM cells are vulnerable to ``1'' to ``0'' bitflips than ``0'' to ``1'' bitflips. Table~\ref{tab:ber_vals} lists the average bitflip count ({across all victim rows}) of both ``1'' to ``0'' than ``0'' to ``1'' bitflips for all the DRAM {chips} we test.

\begin{table}[h]
\centering
\caption{Average bitflip count ({across all victim rows}) of both ``0'' to ``1'' and ``1'' to ``0'' bitflips (Double-Sided RowHammer).}
\label{tab:ber_vals}
\resizebox{\columnwidth}{!}{%
\begin{tabular}{@{}c|c|c||cc|c|c@{}}
\toprule
\multirow{2}{*}{\textbf{Mfr.}} &
  \multirow{2}{*}{\textbf{Die Density}} &
  \multirow{2}{*}{\textbf{Die Revision}} &
  \multicolumn{2}{c|}{\textbf{\begin{tabular}[c]{@{}c@{}}Average Bitflip Count\\ (Across All Rows)\end{tabular}}} &
  \multirow{2}{*}{\textbf{Difference}} &
  \multirow{2}{*}{\textbf{\begin{tabular}[c]{@{}c@{}}Avg. Difference\\ (Geo. Mean)\end{tabular}}} \\
  &      &   & 0 to 1  & 1 to 0  &         &                            \\ \midrule
S & 8Gb  & B & 1769  & 3162   & 78.7\% & \multirow{12}{*}{105.1\%} \\
S & 8Gb  & D & 8617  & 18803  & 118.2\% &                            \\
S & 8Gb  & E & 10414 & 25722  & 147.0\% &                            \\
S & 16Gb & M & 6235  & 13631  & 118.6\% &                            \\
S & 16Gb & A & 6070  & 13833  & 127.9\% &                            \\
S & 16Gb & B & 2496  & 5564   & 122.8\% &                            \\
S & 16Gb & C & 9621  & 23849  & 147.9\% &                            \\ \cmidrule(r){1-6}
H & 8Gb  & C & 2461  & 5417   & 120.1\% &                            \\
H & 8Gb  & D & 2619  & 5226   & 99.5\% &                            \\
H & 16Gb & A & 2295  & 4807   & 109.4\% &                            \\
H & 16Gb & C & 3586  & 6320   & 76.2\% &                            \\ \cmidrule(r){1-6}
M & 8Gb  & E & 3555  & 4593   & 29.2\% &                            \\ \bottomrule
\end{tabular}%
}
\end{table}

We {derive} the following key takeaway.

\takeaway{For Double-Sided RowHammer, significantly more DRAM cells are vulnerable to the error mechanism for ``1'' to ``0'' bitflips than ``0'' to ``1'' bitflips, when the aggressor rows are hammered enough times.}

To further study the relationship between the number of ``1'' to ``0'' and ``0'' to ``1'' bitflips for double-sided RowHammer, we {identify exactly when} the number of ``1'' to ``0'' bitflips start to exceed the number of ``0'' to ``1'' bitflips. {To do so, we gradually increase the aggressor row activation count (with a step size of 1000) from the \hcf of ``0'' to ``1'' bitflips (denoted as {HC\textsubscript{First}\textsubscript{0$\rightarrow$1}}) until we observe more ``1'' to ``0'' bitflips than ``0'' to ``1'' bitflips. We define the aggressor row activation count at this point as {HC\textsubscript{1$\rightarrow$0Exceeds0$\rightarrow$1}}. Table~\ref{tab:sss} compares the average \hcs to the average {HC\textsubscript{First}\textsubscript{0$\rightarrow$1}} for all DRAM chips we test. We observe that, on average, \hcs is $406.5\%$ higher than {HC\textsubscript{First}\textsubscript{0$\rightarrow$1}}.}

\begin{table}[h]
\centering
\caption{Average \hcs compared to {HC\textsubscript{First}\textsubscript{0->1}} (Double-Sided RowHammer).}
\label{tab:sss}
\resizebox{\columnwidth}{!}{%
\begin{tabular}{@{}c|c|c||cc|c|c@{}}
\toprule
\multirow{2}{*}{\textbf{Mfr.}} &
  \multirow{2}{*}{\textbf{Die Density}} &
  \multirow{2}{*}{\textbf{Die Revision}} &
  \multicolumn{2}{c|}{\textbf{Aggr. Row Act. Count}} &
  \multirow{2}{*}{\textbf{Difference}} &
  \multirow{2}{*}{\textbf{\begin{tabular}[c]{@{}c@{}}Avg. Difference\\ (Geo. Mean)\end{tabular}}} \\
  &      &   & {HC\textsubscript{First}\textsubscript{0$\rightarrow$1}} & \hcs  &         &                            \\ \midrule
S & 8 Gb  & B & 43840 & 241740 & 451.4\% & \multirow{12}{*}{406.5\%} \\
S & 8 Gb  & D & 15398 & 63198  & 310.4\% &                            \\
S & 8 Gb  & E & 9684  & 31927  & 229.7\% &                            \\
S & 16 Gb & M & 16732 & 72188  & 331.4\% &                            \\
S & 16 Gb & A & 16981 & 78820  & 364.2\% &                            \\
S & 16 Gb & B & 26415 & 153826 & 482.3\% &                            \\
S & 16 Gb & C & 11355 & 36751  & 223.6\% &                            \\ \cmidrule(r){1-6}
H & 8 Gb  & C & 26500 & 156087 & 489.0\% &                            \\
H & 8 Gb  & D & 22069 & 141656 & 541.9\% &                            \\
H & 16 Gb & A & 29825 & 175674 & 489.0\% &                            \\
H & 16 Gb & C & 18042 & 154951 & 758.8\% &                            \\ \cmidrule(r){1-6}
M & 8 Gb  & E & 44468 & 235454 & 429.5\% &                            \\ \bottomrule
\end{tabular}%
}
\end{table}

\subsection{Bitflip Count with Single-Sided RowPress}
\noindent\textbf{Metric.} To align the experimental characterization results with {Characteristic} 2, we test the maximum number of bitflips that can be induced within the refresh window (i.e., 64ms) for single-sided RowPress for both ``0'' to ``1'' and ``1'' to ``0'' bitflips. 

\noindent\textbf{Methodology.} To cover both the NWL aggressor case and PWL aggressor case {(Section~\ref{sec:rowpressmechanism})}, for each victim row, we perform single-sided RowPress on both its upper {(i.e., \texttt{victim row address + 1}) aggressor row and lower (i.e., \texttt{victim row address - 1})} aggressor row. We keep the aggressor row open for $7.8\mu s$ as 1) it is a {valid} row open time upperbound indicated by the JEDEC DDR4 standard~\cite{jedec2017ddr4}, and 2) it is long enough {for} the RowPress effect {to dominate} the RowHammer effect~\cite{Luo2023RowPress}. {To induce as many RowPress bitflips as possible, we 1) activate the aggressor row as many times as possible within the 64ms refresh window~\cite{jedec2017ddr4} (7500 times),} and 2) {maintain} DRAM temperature at 80$^\circ$C since prior works show that DRAM is much more vulnerable to {single-sided} RowPress at {higher temperatures}~\cite{Luo2023RowPress, Zhou2024Unveiling, Zhou2024Understanding}. We initialize the victim row with either all-zeros (\texttt{0x00}) or all-ones (\texttt{0xFF}), and initialize the aggressor row with the opposite data of the victim row.

\noindent\textbf{Results.}  Figure~\ref{fig:ds_rpber_upper} plots the distribution of the total number of bitflips values per victim row (y-axis) with both ``0'' to ``1'' (i.e., the victim row has data \texttt{0x00}) and ``1'' to ``0'' (i.e., the victim row has data \texttt{0xFF}) bitflips (x-axis) for all DRAM {chips} we test in a box and whiskers plot. Blue represents the upper aggressor row case, and red represents the lower aggressor row case. We make the following observation from the results.
 
\begin{figure}[h]
    \centering
    \includegraphics[width=1.0\linewidth]{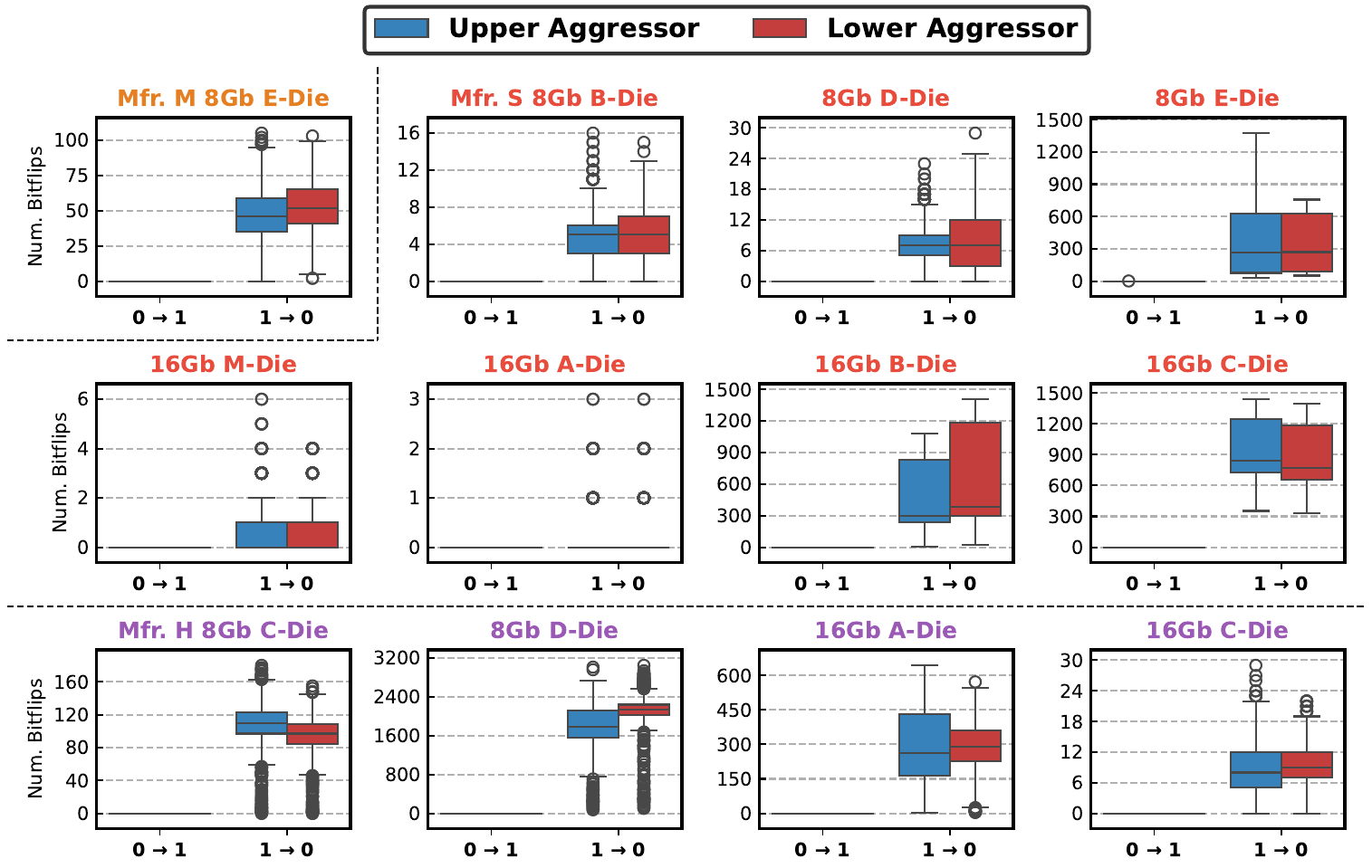}
    \caption{{The distribution of the number of ``0'' to ``1'' and ``1'' to ``0'' bitflips per victim row (Single-Sided RowPress).}}
    \label{fig:ds_rpber_upper}
\end{figure}

\observation{Single-sided RowPress {overwhelmingly} induces ``1'' to ``0'' bitflips.}

We observe that for almost all the DRAM {chips} we test, Single-Sided RowPress induces {only} ``1'' to ``0'' bitflips but not ``0'' to ``1'' {bitflips}. The only exception is Mfr. S 8Gb E-Die where {\emph{only one} victim row has \emph{only one}} ``0'' to ``1'' bitflip. We do not observe {a} significant difference between the upper aggressor row case and the lower aggressor row case{:} on average across all DRAM {chips} we test, the number of bitflips that the upper aggressor induces is {only $3.1\%$ smaller than} that of the lower aggressor row.

\takeaway{For Single-Sided RowPress, the {observed} error mechanism of ``1'' to ``0'' bitflips is much stronger than that of ``0'' to ``1'' bitflips.}

\section{Inconsistencies and Implications}
In this section, we first summarize the key inconsistencies between our experimental characterization results and the device-level read disturbance error mechanism prior works study. Second, we discuss the implications of our findings and propose hypotheses to explain the inconsistencies.
\subsection{Inconsistencies Between Experimental Characterization and Device-Level Studies}
\noindent\textbf{Inconsistency 1.} State-of-the-art device-level error mechanism to explain double-sided RowHammer~\cite{Zhou2023Double} shows that double-sided RowHammer induces {only} ``1'' to ``0'' bitflips (Characteristic 1). However, our experimental characterization results {with real DRAM chips} show that double-sided RowHammer induces \emph{both} ``1'' to ``0'' and ``0'' to ``1'' bitflips.

\noindent\textbf{Inconsistency 2.} State-of-the-art device-level error mechanism to explain double-sided RowHammer~\cite{Zhou2023Double} demonstrates that the consecutive activation of NWL and PWL significantly enhances electron migration into the victim DRAM cell, thus enhancing ``1'' to ``0'' bitflips and mitigating ``0'' to ``1'' bitflips (Characteristic 1). However, our experimental characterization results show that {double-sided RowHammer induces the initial ``0'' to ``1'' bitflips easier (i.e., requiring fewer aggressor row activations) than the initial ``1'' to ``0'' bitflips. This implies that} the {error} mechanism for ``0'' to ``1'' bitflips is stronger than ``1'' to ``0'' bitflips in the most vulnerable cells (i.e., {those} requiring the least number of aggressor row activations to induce bitflips) .

\noindent\textbf{Inconsistency 3.} State-of-the-art device-level error mechanism to explain single-sided {RowPress}~\cite{Zhou2024Unveiling, Zhou2024Understanding} {shows} that {RowPress} should induce both ``1'' to ``0'' and ``0'' to ``1'' bitflips. However, our experimental characterization results show that even with a long aggressor row open time (i.e., $7.8\mu s$), a high aggressor row activation count (i.e., 7500), and using both the NWL and the PWL as the aggressor {rows}, the {overwhelming majority} of single-sided RowPress bitflips {(except for only one)} are ``1'' to ``0'' bitflips.

\subsection{Potential Explanations for the Inconsistencies}
The inconsistencies we identify implies that either 1) existing research on the device-level error mechanisms of DRAM read disturbance are \emph{not comprehensive enough} to cover \emph{all} the major leakage mechanisms, or 2) the state-of-the-art true- and anti-cell reverse engineering technique based on DRAM cell retention failures is incorrect.

Investigating the exact causes of the inconsistencies is outside the scope of this work. However, we still hypothesize several potential reasons that may cause the inconsistencies. First, existing device-level works often make assumptions and simplications when performing {device-level} simulations. For example, prior works that study the trap-assisted electron migration leakage mechanism only focus on a single acceptor-like trap~\cite{yang2019trap, Zhou2023Double, Zhou2024Unveiling, Zhou2024Understanding}, without taking a deeper look into donor-like traps and potential dynamics among multiple trap locations. 

Second, {device-level} simulations usually only model very few isolated {structures and components} (e.g., an active region, two NWLs and two PWLs). This modeling methodology potentially 1) misses key interactions and coupling between multiple devices, and 2) manufacturing variations that could become first-order effects when observed from real-chip experiments. For example, a study that models a whole DRAM row with a statistical modeling of the cell's vulnerability distribution could help with understanding and {explaining} {why Double-Sided RowHammer 1) \emph{first} induces ``0'' to ``1'' bitflips with lower aggressor row activation counts compared to ``1'' to ``0'' bitflips (Takeaway 2), but 2) {induces significantly more ``1'' to ``0'' bitflips} than ``0'' to ``1'' bitflips when the aggressor rows are activated {sufficiently large number} of times (Takeaway 3).} \textcolor{black}{We hypothesize that, similar to DRAM cell retention failure~\cite{Saino2000Impact, Yang2013SuperiorImprovements, Park2015Technology}, there could be two \emph{different} sets of read disturbance leakage mechanisms that affects \emph{different} sets of DRAM cells. For example, while the error mechanism of the ``1'' to ``0'' bitflips could be the major mechanism of Double-Sided RowHammer as prior works study~\cite{Zhou2023Double}, the error mechanism behind the ``0'' to ``1'' bitflips determines the tail distribution of the \hcf (i.e., it affects the most vulnerable DRAM cells).}

Third, the observed bitflips from real-chip experiments are not only affected by the victim cell storage node voltage itself, but also other circuit components and mechanisms like {bitline-bitline} coupling noise~\cite{Min1999Multiple, Li2011DRAMYield, liu2013experimental}, {bitline sense amplifier (BLSA)} offset voltage~\cite{Li2011DRAMYield, Laurent2002SenseAmplifier, Kraus1989Optimized, Lee2010OCBLSA}, etc. For example, if the BLSA design and/or operation causes a significant asymmetry of the signal margin in sensing ``1''s and ``0''s, the real-chip experimental results will be heavily skewed {towards one bitflip direction}.

We hope that our results in this paper provide future works with insights into building a more fundamental and comprehensive understanding of DRAM read disturbance.

\section{Related Work}
To our knowledge, this is the first paper to systematically demonstrate the fundamental inconsistencies of DRAM read disturbance bitflip characteristics between real-chip experimental characterization and proposed device-level error mechanisms. Prior works on experimental characterization~\cite{kim2014flipping, kim2020revisiting, orosa2021deeper, yaglikci2022understanding, Luo2023RowPress, Nam2024DRAMScope, luo2024combined, olgun2024read, olgun2025variable} do not involve alignment with device-level studies. Prior device-level studies~\cite{yang2019trap, walker2021ondramrowhammer, Zhou2023Double, Zhou2024Unveiling, Zhou2024Understanding} do \emph{not} conduct experimental characterization using real DRAM chips that matches the conditions set in their simulations. Earlier work~\cite{park2016experiments, ryu2017overcoming} that conducts both real-chip characterizations and {device-level} simulations do \emph{not} study the double-sided RowHammer and RowPress access patterns.

\section{Conclusion}
 In this paper, we attempt to align and cross-validate the real-chip experimental characterization results and state-of-the-art device-level studies of DRAM read disturbance. Through our experiments, we identify fundamental inconsistencies in the RowHammer and RowPress bitflip directions and access pattern dependence between experimental characterization results and the device-level error mechanisms. 
 
 Based on our results, we hypothesize that either 1) existing research on the device-level error mechanisms of DRAM read disturbance are {not comprehensive enough} to cover {all} the major leakage mechanisms, or 2) the state-of-the-art true- and anti-cell reverse engineering technique based on DRAM cell retention failures is incorrect. We hope our findings {inspire and} enable future works to build a more fundamental and comprehensive understanding of DRAM read disturbance.

\section*{Acknowledgments}
{We thank the anonymous reviewers of {VTS 2025} for feedback. {We thank the} SAFARI Research Group members for
{constructive} feedback and the stimulating intellectual {environment.} We
acknowledge the generous gift funding provided by our industrial partners
({especially} Google, Huawei, Intel, Microsoft), which has been instrumental in
enabling the research we have {been} conducting on read disturbance in DRAM {in
particular and memory systems in
general~\cite{mutlu2023retrospectiveflippingbitsmemory, mutlu2021primer}.} This work was in part
supported by {a} Google Security and Privacy Research Award and the Microsoft
Swiss Joint Research Center.}

\bibliographystyle{unsrt}
\balance
\bibliography{refs}

\end{document}